\documentstyle[aps,prl,epsf,twocolumn]{revtex}
\begin{document}
\def\bea{\begin{eqnarray}}
\def\eea{\end{eqnarray}}
\def\a{\alpha}
\def\d{\delta}
\def\p{\partial} 
\def\nn{\nonumber}
\def\r{\rho}
\def\rv{\bar{r}}
\def\la{\langle}
\def\ra{\rangle}
\def\e{\epsilon}
\def\o{\omega}
\def\n{\eta}
\def\g{\gamma}
\def\break#1{\pagebreak \vspace*{#1}}
\def\f{\frac}
\draft
\title{Elasticity of Semi-flexible Polymers}
\author{Joseph Samuel$^{1}$\cite{SAM}and 
Supurna Sinha$^{2}$\cite{SUP}}
\address{Raman Research Institute,
Bangalore 560080,India\\}
\maketitle
\widetext
\begin{abstract}
We present an exact solution of the Worm-Like Chain (WLC) model for 
semi-flexible polymers valid over {\it the entire range of polymer lengths}.
Our results
are in excellent agreement with recent computer simulations and 
reproduces important qualitatively interesting features observed in simulations
of polymers of intermediate lengths. We also make a number of predictions that 
can be tested in a variety of concrete experimental realizations. The expected
level of finite size fluctuations in force-extension curves is also 
estimated. 
This study is relevant to mechanical properties of
biological molecules.
\end{abstract}

\narrowtext
Many biologically important molecules, like DNA and Actin,
are semi-flexible polymers\cite{Marko}. In recent years
there have been experiments\cite{bust} which pull and stretch single
molecules to 
measure elastic 
properties.
 For instance, one can study the
``equation of state'' of a 
semi-flexible polymer by measuring its extension\cite{shiva} as a 
function of applied
force. Alternatively one can tag the ends with fluorescent dye\cite{ott}
and
determine the distribution
of end-to-end distances. Such studies reveal a wealth of information 
about the mechanical 
properties of semi-flexible polymers, which is of clear biological importance. 

Till a few years ago, studies of polymer molecules such as
DNA were limited to samples
containing large numbers of molecules \cite{bhat}. This made it hard to probe the
elastic properties of individual DNA which are of
vital importance to biological
processes such as protein-induced DNA bending \cite{bust,bouchiat}.
It is only quite recently, due to advances in technology
that single molecule studies became feasible.
In order to correctly interpret single molecule
experiments which are
now being performed, a good theoretical understanding of
semi-flexible polymers is essential. Quite apart from the biological
interest, semi-flexible polymers are  
of interest to physicists \cite{doi,Gobush}. This paper is devoted
to understanding the equilibrium statistical mechanics of 
single
semi-flexible polymers.

Statistical mechanics of a single polymer molecule 
is dominated by fluctuations because it is a system of finite size. It is 
only in the thermodynamic limit of extremely long polymers that these
fluctuations 
about the mean die out. Due to the dominance
of fluctuations, the experimentally measured mean values for a semi-flexible 
polymer crucially depend on the precise choice of the ensemble. For instance,
one gets qualitatively distinct features in force-extension curves depending 
on whether the force or the extension is held constant in an 
experimental setup \cite{dabhi,kreuz}.

The most popular theoretical model for understanding semi-flexible
polymers
is the Worm-Like Chain (WLC)\cite{kratky}, which ignores
self-avoidance and models the polymer
as a framed
space curve of fixed total length $L$ with an energy cost for bending
and twisting. 
\break{1.2in}
In order to interpret the
experimental data, it would be useful to have a clear and
complete understanding of the predictions of the 
WLC model. Such an understanding would reveal the strengths and
deficiencies of the model in describing real 
polymers and could be used to improve the model. 
There do exist partial results \cite{wilhelm}  
on the statistical mechanics
of the WLC
model: some theoretical studies \cite{Marko} 
investigate 
the 
flexible limit of very long polymers (long 
compared to
the 
persistence length;$L>>L_p$). However 
experimental interest
is {\it not confined
to very long polymers}.
For example, experiments on Actin \cite{ott}
deal with polymers of length $L=30\mu m$, which is only about
twice the
measured persistence length of $L_p=16.7 \mu m$.
There is also theoretical work\cite{wilhelm} on 
extremely short polymers.
There is a clear gap in the present
understanding of the WLC model for polymers of intermediate length. 
Our purpose in this paper is to fill this gap. We present a solution 
of the WLC model and 
describe the equilibrium elastic properties expected from the model. 
Our solution 
is exact
in the sense that the elastic properties can be determined to
any desired accuracy.  
The main results of 
this paper are contained in the figures which show the
force-extension 
relation and end-to-end distance distributions predicted by the
WLC model. 
These predictions 
agree well with two independent computer
simulations \cite{wilhelm,dabhi} (Figs.3 and 4).

{\it WLC model with pure bend:} A configuration ${\cal C}$ of the polymer
is
described by a  space curve ${\vec x}(s)$, with $s$ the arc-length
parameter ($0\le s \le L$) ranging from $0$ to $L$, the contour 
length of the 
polymer. 
The tangent vector ${\hat t}=d{\vec x}/ds$  to the curve is a unit vector
\begin{equation}
{\hat t}.{\hat t}=1
\label{unit}
\end{equation}
and the curvature of the polymer is given by $\kappa=|d{\hat t}/ds|$.
We will suppose that one end of the polymer is tethered to the
origin (${\vec x}(0)=0$) and the other
end ${\vec x}(L)={\vec r}$ is tagged.
As the polymer configuration changes with thermal agitation, the
location ${\vec r}$ of its tagged end fluctuates.
The quantity we wish to compute is $Q({\vec r})$, which is the
probability distribution for the location ${\vec r}$ of the
tagged end \cite{dabhi}.
If the tagged end is pulled from ${\vec r}$ to ${\vec r}+d{\vec r}$, 
$Q({\vec r})$ changes and consequently, the free
energy. This implies that a force is needed to
stretch the polymer. Thus $Q({\vec r})$ is directly related to
the force-extension relation of the polymer.
To compute $Q({\vec r})$ we need to sum over all polymer
configurations
${\cal C}$ which end at ${\vec r}$, with a Boltzmann 
weight: $Z=\Sigma_{\cal C} \exp(-{\cal E}[{\cal C}]/k_BT),$
where the energy ${\cal E}$ associated with 
a configuration ${\cal C}$ is ${\cal E}({\cal C})= \frac{1}{2}A\int_0^L ds
\kappa^2$
and $A$ is the {\it bending modulus}.
This is  a standard counting problem in statistical mechanics
and can be naturally addressed in the language
of path integration\cite{Schulman}.
However, not much progress has been made because of the difficulty
\cite{dave,footnoted} presented by the inextensibility constraint
(\ref{unit}). 
The key to circumventing
this difficulty is to consider \cite{Marko,BM} 
Brownian motion in the
space of {\it
tangent vectors} (${\hat t}$)
rather than (as is customary for flexible
polymers) position vectors ${\vec x}$. The tangent vectors form a unit
sphere (See Eq.(\ref{unit})) and the problem reduces
to studying Brownian motion on the unit sphere, which can be handled
by standard operator techniques familiar
from quantum mechanics.

Let us suppose to begin with that the initial 
(${\hat t}_A=\frac{d{\vec x}}{d s}|_{s=0}$)
and final (${\hat t}_B=\frac{d{\vec x}}{d s}|_{s=L}$) tangent 
vectors are held fixed. $Q({\vec r})$ has the path integral representation
\begin{equation}
{\cal N}                                                       
\int{\cal D}[{\hat t}(s)] e^{-1/k_BT[A/2\int_0^L (d{\hat t}/ds)^{2}
ds]}\delta^3({\vec r}-\int_0^L{\hat t}ds)
\label{pathQ}
\end{equation}
where ${\cal N}$ is a normalisation constant. Instead of $Q({\vec r})$
we focus on the quantity $P(z)=\int d{\vec r}Q({\vec r}) \delta(r_3-z)$,
which is $Q({\vec r})$ integrated over a plane of constant $z$.
Note that $P(z)$ and $Q({\vec r})$ vanish when the modulus of their
arguments exceeds $L$. The generating function of $P(z)$ is
defined as 
${\tilde P}(f)=\int_{-L}^{L}dz \,e^{fz/L_p} P(z)$,
where $L_p=A/k_BT$.
Performing the elementary integrations involving $\delta$- functions
we find that
${\tilde P}(f)$ can be expressed as $Z(f)/Z(0)$, where $Z(f)$
has the path integral representation 
\begin{equation}
Z(f)={\cal N}\int{\cal D}[{\hat t}(s)] e^{-L_p/2[\int_0^L(d{\hat
t}/ds)^{2}ds]}e^{[f/L_p\int_0^L{\hat t}_zds]}
\label{path}
\end{equation}
Making the change of variable $\tau= s/L_p$, we arrive
at the expression
\begin{equation}
Z(f)={\cal N}\int{\cal D}[{\hat t}(\tau)] e^{-\int_0^\beta d\tau[1/2(d{\hat t}/d\tau)^2
-f{\hat t}_z]}
\label{path2}
\end{equation}
where $\beta =L/L_p$.
Eq. (\ref{path2}) can be interpreted as the path integral
representation
for the kernel of a {\it quantum} particle on the
surface of a sphere at inverse temperature $\beta$. 
Thus we can express $Z(f)$ as the quantum amplitude to go from
an initial tangent vector $\hat t_A$ to a final tangent vector $\hat t_B$ 
in imaginary time $\beta$ in the presence of an
external potential $-f \cos\theta$:
\begin{equation}
Z(f)=\sum_n e^{-[\beta E_n]}\psi_n^{*}(\hat t_A)\psi_n(\hat t_B).
\label{operz}
\end{equation}
Here $\{\psi_n(\hat t)\}$,
is a complete set of normalized eigenstates of the
Hamiltonian ${\hat H} = -\frac{\nabla^2}{2} -f \cos\theta $ and
$E_n$ are the corresponding eigenvalues.

In this paper we focus on the situation where the 
boundary tangent vectors are unconstrained, {\it i.e} they are integrated
over with uniform weight.
For free boundary conditions the situation is spherically
symmetric and $Q$ depends only on $r=|r|$ and not
on ${\vec r}$ and we can write $Q(r)$. The probability distribution
for the end to end distance $r$ is given by $S(r)=4\pi r^2 Q(r)$,
as can be seen by integrating $Q({\vec r})$ over a sphere of radius $r$. 
As we did above, we can integrate $Q({\vec r})$ over
a plane of fixed $z$:$P(z)=\int d{\vec r}Q({\vec r}) \delta(r_3-z)$.
$P(z)$ is the probability distribution for the $z$ co-ordinate of the
tagged end. Both $S(r)$ and $P(z)$ are experimentally accessible
quantities and they are integrals of the spherically symmetric
function $Q({\vec r})$ 
over two dimensional surfaces. 
Using tomographic techniques (reconstruction
of a function from a knowledge of its integral over two dimensional slices)
one can deduce\cite{foottom} the relation $S(r)=-2 rdP(r)/dr$, where $P(r)$ is 
$P(z)$ with its argument replaced by $r$. 

For free boundary conditions (\ref{operz}) can be written as a``vacuum
persistence amplitude''
\begin{equation}
Z(f) = < 0| \exp - \beta H_f |0 >
\label{part}
\end{equation}
where $H_f = -\frac{1}{2} \nabla^2 -f \cos\theta$ is the Hamiltonian
of the rigid rotor \cite{Marko} in a potential and
$|0>$ is the ground
state of the free Hamiltonian $H_0 = -\frac{1}{2} \nabla^2$. 
\vbox{
\epsfxsize=6.0cm
\epsfysize=6.0cm
\epsffile{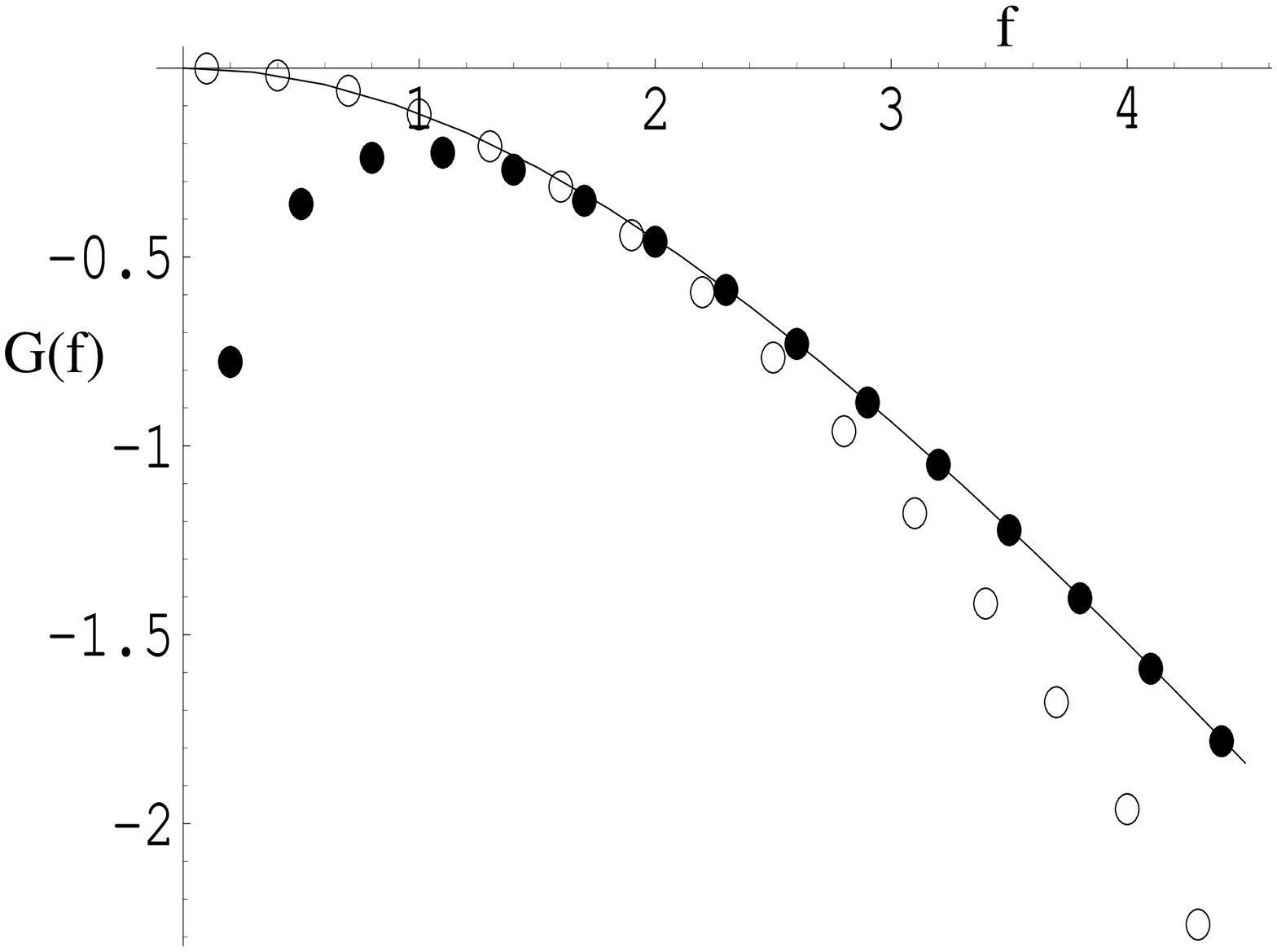}
\begin{figure}
\caption{$G(f)$ as a function of $f$ for
$\beta =L/L_p=1$. Also plotted are approximate analytic forms
valid in the small (open circles) and large (filled circles) force
regimes.}
\label{Gf}
\end{figure}}
 
By choosing a basis in which
$H_0$ is diagonal we find that 
$H$ is a symmetric tridiagonal matrix with diagonal
elements
$H_{l\,\,l}={l(l+1)}/{2}$
and superdiagonal  elements 
$H_{l\,\,l+1}=f (l+1)\sqrt{1/((2l+1)(2l+3))}$. 
Upto this point the treatment is completely analytical. To evaluate 
Eq. (\ref{part}) we need to use numerical methods. 
$H_f$ is really 
an infinite matrix, but we truncate it to $N X N$ size, numerically 
evaluate it (using Mathematica\cite{math}) and adjust the cutoff $N$ until
the answer stabilizes to desired accuracy.  
From this we deduce all the properties of the model, to an
accuracy limited only by computational power.
The form of $G(f)= -1/\beta\log{Z(f)}$
is shown in Fig. 1 along with physically motivated 
approximate
analytical forms valid in the small ( $G(f)= C_1(\beta) f^2$) and large 
($G(f)=-f+\sqrt{f}-\log{f}/(2\beta)+C_2(\beta)$) force regimes.
More terms can be computed, but this already gives a fair fit.

From $\tilde{P}(f)$ (which is equal to $Z(f)$ since $Z(0)=1$),
it is possible to compute $P(z)$ by performing
the inverse Laplace transform. (Numerically it is more convenient to
use the inverse Fourier transform by going to imaginary $f$).
The results are shown in Figure 2. For convenience, we set $L_p=1$ so that
$\beta=L$ and 
plot all figures in terms of scaled variables,
$\zeta=z/\beta$, $\rho=r/\beta$.
\vbox{
\epsfxsize=6.0cm
\epsfysize=6.0cm
\epsffile{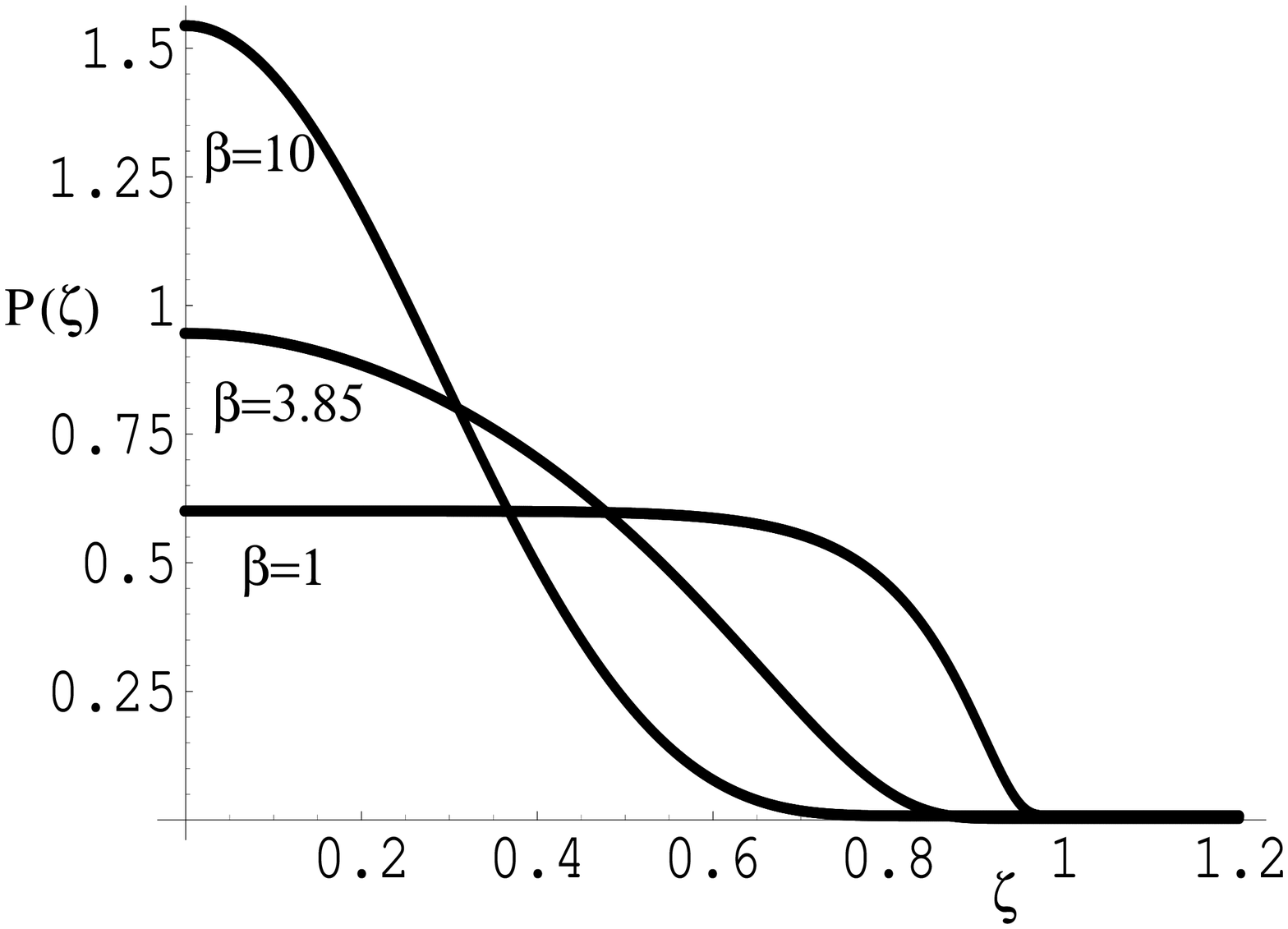}
\begin{figure}
\caption{
The distribution $P(\zeta)$ of scaled extension $\zeta=z/\beta$
for $\beta = L/L_p$ equal
to 1, 3.85 and 10.}
\label{Pz}
\end{figure}}
\vbox{
\epsfxsize=6.0cm
\epsfysize=6.0cm
\epsffile{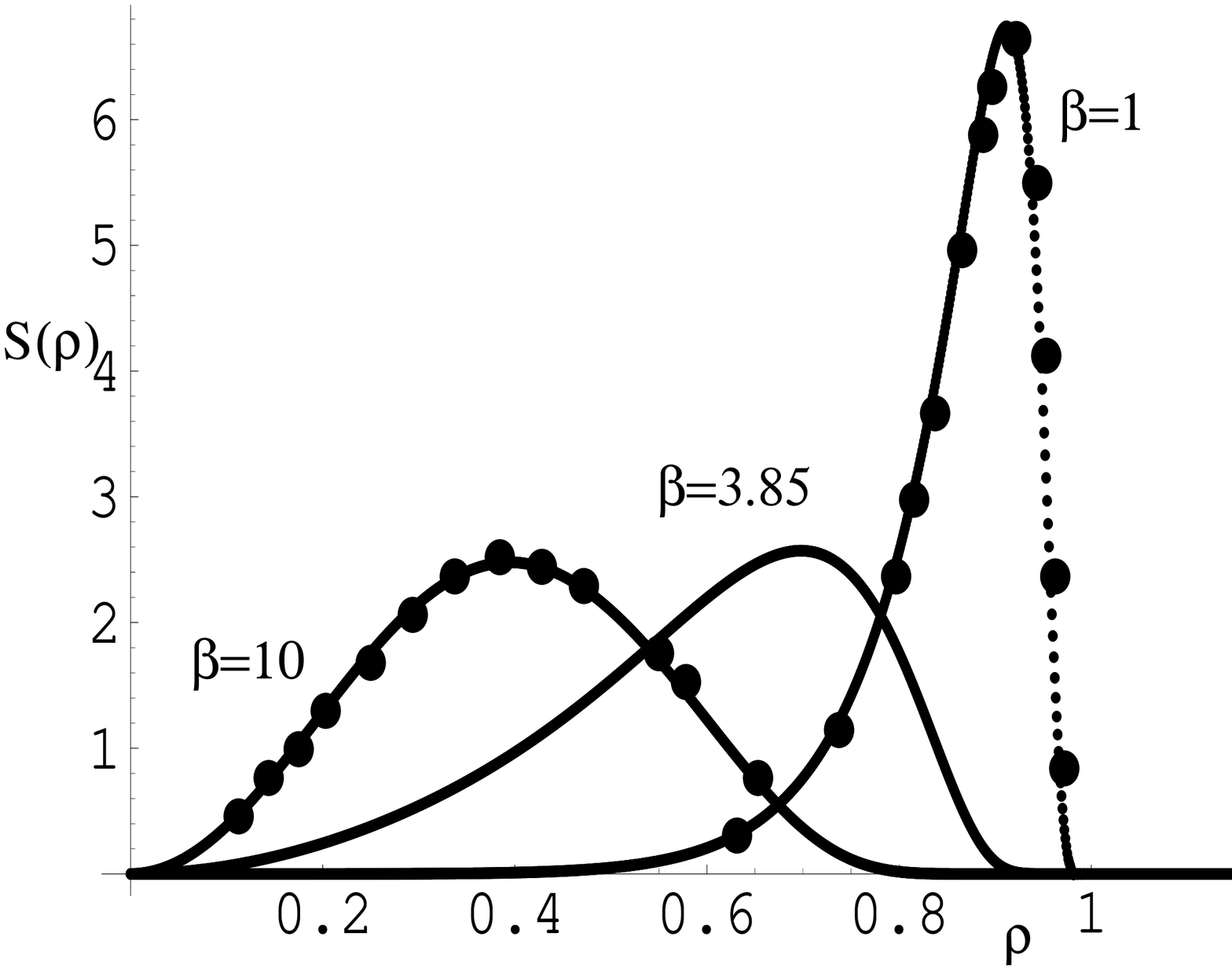}
\begin{figure}
\caption{The distribution $S(\rho)$ of scaled
end-to-end distances $\rho=r/\beta$ for
$\beta = L/L_p$ equal to 1, 3.85 and 10. Dots show simulation
data taken from Ref.[9].}
\label{Sr}
\end{figure}}
From the relation $S(\rho) = -2\rho\frac{d}{d\rho} P(\rho)$
we compute the
distribution of end-to-end distance. These are displayed in Fig. 3.
We have checked that these graphs quantitatively agree (to
within the errors of the simulation data) with
the published plots of \cite{wilhelm}.
Notice that $P(\zeta)$ and $S(\rho)$ both have a single maximum and
the corresponding free energies have a single minimum.
However,
for a range of  $\beta$ near 3.8, $Q(\rho)$ develops a double humped form,
reflecting the existence of two stable free energy minima resulting in
a ``first order transition'', where the quotes signify that this
is not a true phase transition due to finite size effects.
This feature was first noticed
\cite{dabhi} in computer simulations of the WLC model.
Our theoretical work confirms the results of
simulations presented in \cite{dabhi}.
The form of $Q(\rho)$ is plotted in Fig.4
along with the results of computer simulations from \cite{dabhi}.
\vbox{
\epsfxsize=6.0cm
\epsfysize=6.0cm
\epsffile{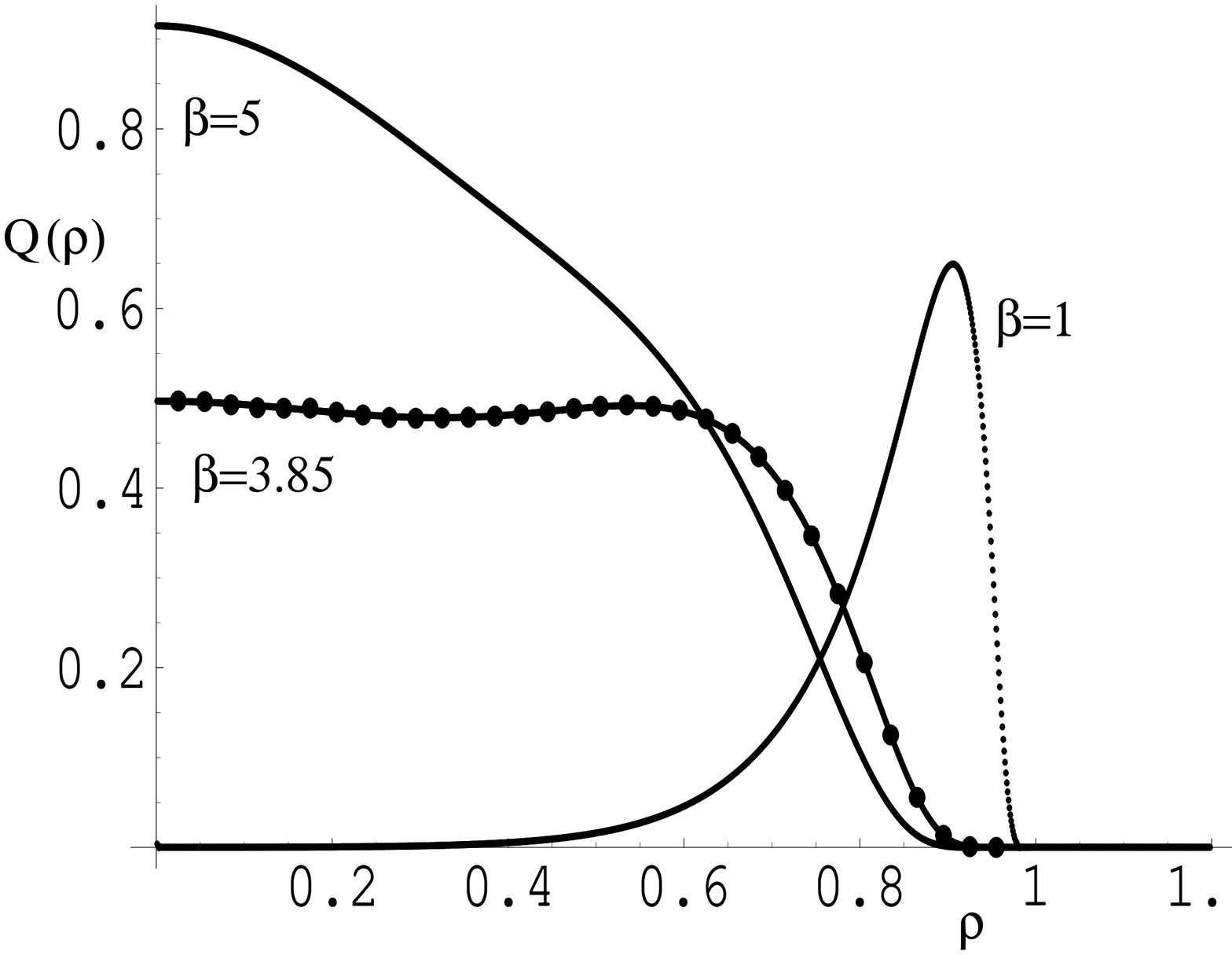}
\begin{figure}
\caption{The function $Q(\rho)$ for $\beta=L/L_p = 1,3.85$
and $5$. Results of a simulation from Ref.[10] are also plotted
on the curve for $\beta=3.85$.}
\label{Qr}
\end{figure}}
\vbox{
\epsfxsize=6.0cm
\epsfysize=6.0cm
\epsffile{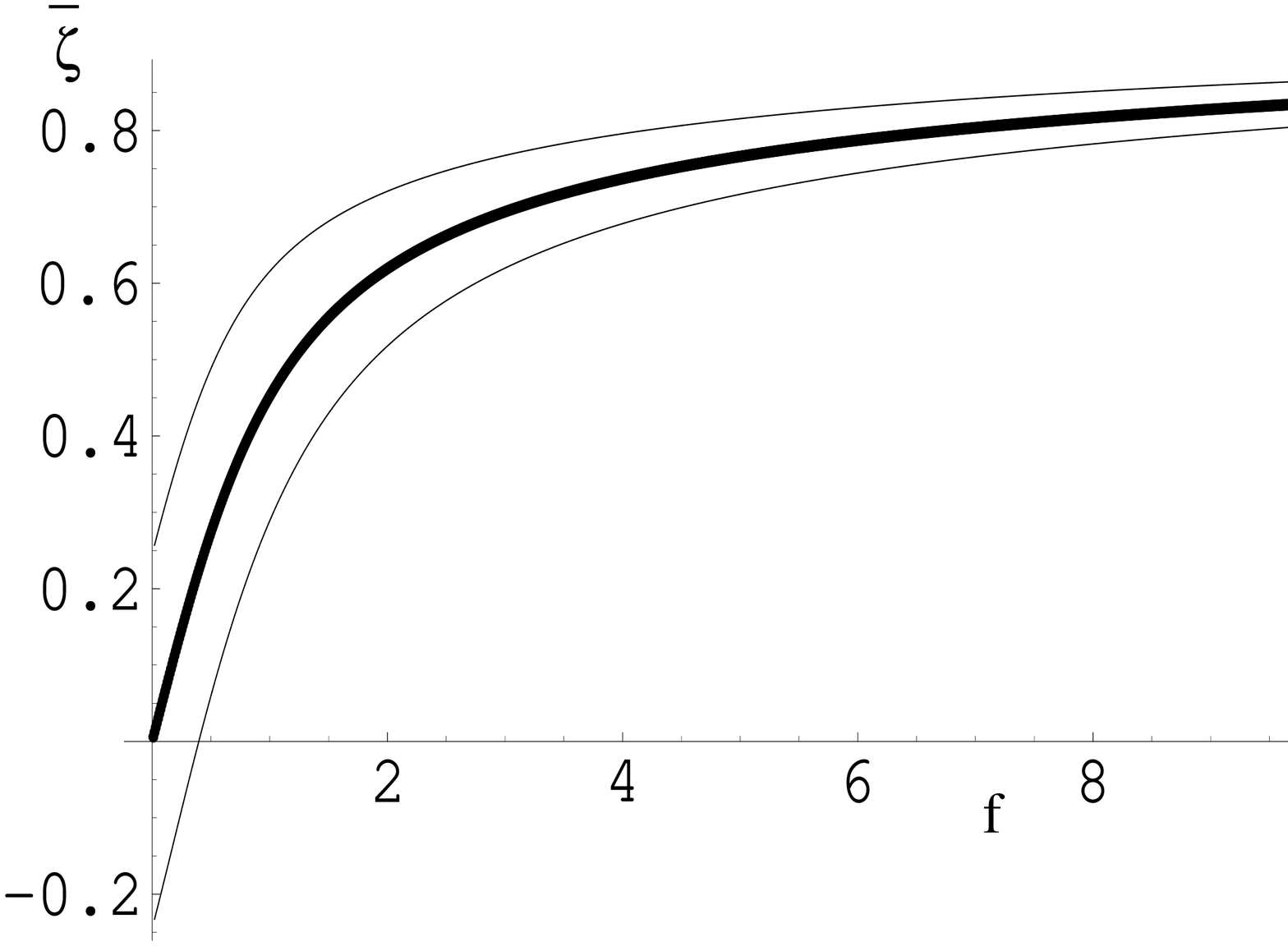}
\begin{figure}
\caption{The mean extension ${\bar \zeta}$ as a function of
$f$ (thick line) for $\beta=L/L_p=10$. Also shown on either side are
the root mean
square fluctuations (thin lines) of the extension about its mean value.}
\label{Fext}
\end{figure}}
A property of direct experimental interest is the force-extension
relation (FER).
We work in the constant $f$ ensemble and in Fig. 5 plot the scaled
mean extension $\overline{\zeta}(f)$ (defined
by ${\overline{\zeta}}=-\partial G(f)/\partial f$).

Since we are dealing with a system of finite size, we expect that the 
extension $\zeta$ will fluctuate about its mean value ${\overline{\zeta}}$.
The theoretically expected
root mean square value of these fluctuations 
$\Delta \zeta=\sqrt{1/\beta {\partial \overline{\zeta}}/{\partial f}}$
of $\zeta$ is shown in Fig. 5. These fluctuations clearly vanish in the limit of infinitely long polymers \cite{Marko}.

Although the WLC model has been known for around
fifty years\cite{kratky}, there does not appear to be a closed form
analytic solution
in terms of elementary functions. This paper presents a numerical solution
to the WLC, which, given the power of modern personal computers is as useful as
an exact analytic form. Using the techniques outlined in this paper one
could work out the predictions of the WLC model
to any desired accuracy, for example, experimental accuracy.
Our work provides predictions
of force-extension curves for {\it all}
lengths
which can be tested against experiments.
All the quantities computed
here $(Q(\rho), P(\zeta), S(\rho)$, $\tilde{P}(f)$, $G(f)$ and FER) are
experimentally measurable. Here we briefly go over the pertinent experimental
realizations of some of these quantities.
(a) Measurement of $Q(\rho)$: One can measure $Q(\rho)$ by attaching a bead
to one end of the molecule and confining the bead in a stiff optical trap and by
recording the distribution of location of the other end (tagged with dye) by
means of a CCD camera.
(b) Measurement of $P(\zeta)$: $P(\zeta)$ can be measured by recording the
location of the free end of the molecule on a given $\zeta$ plane and
focussing all the light from the particular $\zeta$ plane by using a confocal
microscope.

In this paper we have used free boundary conditions for the tangent vectors.
Other boundary conditions can also be handled
as explained in \cite{mmg}. The boundary conditions depend on the
particular experimental
setup. For example, if the tangent vectors at the ends are held fixed one
needs to use $\delta$ function weights rather than uniform ones.
Choice of theoretical weights consistent with experimental boundary
conditions is particularly crucial in the context of short polymers.

The FERs predicted by the model depend on the ensemble in which the calculation
is done. Depending on the experimental situation one should use an ensemble
in which one of $z, r, \vec r$ or their conjugate forces $f, f_r, \vec f$
is held constant. As an example we display in Fig.5 the FER in
the constant $f$ ensemble. The force extension relations in this ensemble
are monotonic for {\it all} values of $\beta$. (In contrast, the FER
in the constant $\rho$ ensemble is non-monotonic \cite{dabhi} in the
$\beta$ range where the function $Q(\rho)$ is double humped.)
As mentioned earlier, since we are dealing with a finite
system, which is not near the thermodynamic limit, equivalence 
between conjugate ensembles is not assured \cite{kreuz}. This is  due
to fluctuations 
around mean values which are not negligible 
for short polymers. Our analysis, being exact, correctly takes into
account such finite size effects. 
The pure bend WLC model has a single parameter $\beta=L/L_p$ and predicts
not only a force extension relation but also the 
amount of theoretically expected noise on this curve. 
In an experiment one can expect to see this noise over and
above any instrumental noise that may be present in the system.

Self-avoidance is a feature present in real polymers, which has not been 
taken into account in this analysis. 
Such effects are more important for flexible polymers. Self-avoidance is 
difficult to handle analytically and is one of the important directions
for future work. 
We hope that the results of this paper will stimulate a detailed and
quantitative comparison between the predictions of the WLC model and 
experiments, 
and lead to an improved understanding of the elasticity of
semi-flexible polymers. 

{\it Note added:} After this manuscript was submitted for publication,
we learned of closely related work by Stepanow and Sch{\"u}tz
\cite{Step}.

{\it Acknowledgements:}
It is a pleasure to thank Abhishek Dhar and V. A. Raghunathan for their
critical comments and 
Y. Hatwalne, M. Rao, T. Roopa and
G. V. Shivashankar for discussions.


\begin{references}
\bibitem{footnoted}The inextensibility constraint has been
enforced in an average or mean-field sense in \cite{dave}.
\bibitem{foottom}
We deduce the relation by using the fact that the Fourier transform
${\tilde Q(\vec k)}$
of $Q(\vec r)$ is related to $\tilde P(k)$, the Fourier transform of
$P(z)$ via
$\tilde P({k})=(2\pi)^2\tilde Q(0,0,|k|)$. Thus we get $Q(\vec r) =
\int {{\tilde Q}(0,0,|k|)
e^{-i{\vec k}.{\vec r}}d {\vec k}} $.
Performing the angular integration yields the result.
\end{references}
\end{document}